\RequirePackage{lineno}
\documentclass[twocolumn,showpacs,aps,prd]{revtex4-1}
\usepackage{graphicx}
\usepackage{epsfig,graphics,subfigure,psfrag,amsmath,amssymb}
\usepackage{dcolumn}
\usepackage{bm}
\usepackage{rotating}
\usepackage{multirow}
\usepackage{epsfig}
\usepackage{float}
\usepackage{mathrsfs}
\usepackage{booktabs}
\usepackage{subfigure}
\usepackage{caption}
\usepackage{color}
\begin{document}
\normalsize
\parskip=5pt plus 1pt minus 1pt

\title{\boldmath Search
for the isospin violating decay $Y(4260)\rightarrow J/\psi \eta
\pi^{0}$}

\newcommand{\tabincell}[2]{\begin{tabular}{@{}#1@{}}#2\end{tabular}}

\author{
  \small
      M.~Ablikim$^{1}$, M.~N.~Achasov$^{9,f}$, X.~C.~Ai$^{1}$,
      O.~Albayrak$^{5}$, M.~Albrecht$^{4}$, D.~J.~Ambrose$^{44}$,
      A.~Amoroso$^{48A,48C}$, F.~F.~An$^{1}$, Q.~An$^{45,a}$,
      J.~Z.~Bai$^{1}$
% ...
      \\
      \vspace{0.2cm}
      (BESIII Collaboration)\\
      \vspace{0.2cm} {\it
        $^{1}$ Institute of High Energy Physics, Beijing 100049, People's Republic of China\\
% ...
$^{4}$ Bochum Ruhr-University, D-44780 Bochum, Germany\\
        $^{5}$ Carnegie Mellon University, Pittsburgh, Pennsylvania 15213, USA\\
% ...
$^{9}$ G.I. Budker Institute of Nuclear Physics SB RAS (BINP), Novosibirsk 630090, Russia\\
% ...
$^{44}$ University of Rochester, Rochester, New York 14627, USA\\
        $^{45}$ University of Science and Technology of China, Hefei 230026, People's Republic of China\\
% ...
$^{48}$ (A)University of Turin, I-10125, Turin, Italy; (B)University of Eastern Piedmont, I-15121, Alessandria, Italy; (C)INFN, I-10125, Turin, Italy\\
% ...
$^{a}$ Also at State Key Laboratory of Particle Detection and Electronics, Beijing 100049, Hefei 230026, People's Republic of China\\
% ...
$^{f}$ Also at the Novosibirsk State University, Novosibirsk, 630090, Russia\\
% ...
      }
}

\date{\today}

\begin{abstract}
Using data samples collected at center-of-mass energies of $\sqrt{s}$
= 4.009, 4.226, 4.257, 4.358, 4.416, and 4.599 GeV with the BESIII detector operating at the BEPCII storage ring, we search for the
isospin violating decay $Y(4260)\rightarrow J/\psi \eta \pi^{0}$. No
signal is observed, and upper limits on the cross section
$\sigma(e^{+}e^{-}\rightarrow J/\psi \eta \pi^{0})$ at the 90\%
confidence level are determined to be 3.6, 1.7, 2.4, 1.4, 0.9, and 1.9
pb, respectively.
\end{abstract}

\maketitle

\section{Introduction}
The $Y(4260)$ charmoniumlike state was first observed in its decay to
$\pi^{+} \pi^{-} J/\psi$~\cite{bib1} and has a small coupling to open
charm decay modes~\cite{bib2}. $Y(4260)$ is a vector ($J^{PC} = 1^{--}$) state that is
only barely observable as an s-channel resonance in $e^{+}e^{-}$
collisions and that appears at an energy where no conventional
charmonium state is expected. Since its discovery, many theoretical
studies have been carried out considering the $Y(4260)$ as a
tetraquark state~\cite{bib3}, $D_{1}D$ or $D_{0}D^{*}$ hadronic
molecule~\cite{bib4}, hybrid charmonium~\cite{bib5}, baryonium state
\cite{bib6}, etc.

Recently, in the study of $Y(4260)\rightarrow \pi^{+} \pi^{-}
J/\psi$, a charged charmoniumlike structure, the $Z_{c}(3900)^{\pm}$, was observed in the
$\pi^{\pm}J/\psi$ invariant mass spectrum by the BESIII~\cite{bib7} and Belle
experiments~\cite{bib8} and confirmed shortly thereafter with CLEO-c data~\cite{bib9}. In the
molecule model~\cite{bib10}, the $Y(4260)$ is
proposed to have a large $D_{1}\bar{D}$ component, while
$Z_{c}(3900)^{\pm}$ has a $D\bar{D}^{*}$ component.

BESIII recently reported
the observation of $e^{+}e^{-}\rightarrow \gamma X(3872)\rightarrow
\gamma \pi^{+} \pi^{-} J/\psi$~\cite{bib11}. The cross section
measurements strongly support the existence of the radiative
transition $Y(4260)\rightarrow \gamma X(3872)$. One significant
feature of the $X(3872)$ that differs from conventional charmonium is
that the decay branching fraction of $X(3872)$ to $\pi^{+} \pi^{-} \pi^{0}
J/\psi$ is comparable to $\pi^{+} \pi^{-}
J/\psi$~\cite{bib12,bib13}, so the isospin violating process occurs on
a large scale.

Isospin violating decays can be used to probe the nature of heavy
quarkonium. The hadro-charmonium model~\cite{bib14} and tetraquark
models~\cite{bib15,bib16} predict that the reaction
$\Upsilon(5S)\rightarrow \eta \pi^{0} +$
bottomonium should be observable. The tetraquark model~\cite{bib17} also predicts
that $Z^{0}_{c}$ can be produced in $Y(4260) \rightarrow J/\psi \eta
\pi^{0}$ with $Z^{0}_{c}$ decaying into $J/\psi\pi^{0}$ and possibly
$J/\psi\eta$ in the presence of sizable isospin violation. The molecular model~\cite{bib18} predicts a peak
in the cross section of $Y(4260) \rightarrow J/\psi \eta \pi^{0}$ at the
$D_{1}\bar {D}$ threshold and a narrow peak in the $J/\psi\eta$
invariant mass spectrum at the $D\bar {D^{*}}$ threshold.

In this paper, we present results on a search for the isospin
violating decay $Y(4260)\rightarrow J/\psi \eta \pi^{0}$, with $J/\psi
\rightarrow e^{+}e^{-}/\mu^{+}\mu^{-}$, $\pi^{0} \rightarrow \gamma\gamma$, and $\eta\rightarrow
\gamma\gamma$ (the other decay modes of $\eta$ are not used due to much lower detection efficiency
and branching fraction), based on
$e^{+}e^{-}$ annihilation data collected with the BESIII detector
operating at the BEPCII storage ring~\cite{bib19} at center-of-mass
energies of $\sqrt{s}$ = 4.009, 4.226, 4.257, 4.358, 4.416, and 4.599 GeV.

\section{BESIII detector and Monte Carlo Simulation}
The BESIII detector, described in detail in Ref.~\cite{bib19}, has a
geometrical acceptance of 93\% of 4${\pi}$. A small-cell helium-based
main drift chamber (MDC) provides a charged particle momentum
resolution of 0.5\% at 1 GeV/$c$ in a 1 T magnetic field and supplies
energy-loss ($dE/dx$) measurements with a resolution of 6\%
for minimum-ionizing pions. The electromagnetic calorimeter
(EMC) measures photon energies with a resolution of 2.5\% (5\%) at 1.0
GeV in the barrel (end caps). Particle identification is provided
by a time-of-flight system with a time resolution of 80 ps (110
ps) for the barrel (end caps). The muon system (MUC), located in the iron
flux return yoke of the magnet, provides 2 cm position resolution and
detects muon tracks with momentum greater than 0.5 GeV/$c$.

The {\footnotesize GEANT}4-based~\cite{bib20} Monte Carlo (MC)
simulation software {\footnotesize BOOST}~\cite{bib21} includes the
geometric description of the BESIII detector and a simulation of the
detector response. It is used to optimize event selection criteria,
estimate backgrounds, and evaluate the detection efficiency. For each
energy point, we generate large signal MC samples of $e^{+}e^{-} \rightarrow
J/\psi\eta\pi^{0}$, $J/\psi \rightarrow e^{+}e^{-}/\mu^{+}\mu^{-}$,
$\eta\rightarrow \gamma\gamma$, and $\pi^{0} \rightarrow \gamma\gamma$
uniformly in phase space. Effects of initial state radiation (ISR) are
simulated with {\footnotesize KKMC}~\cite{bib22}, where the Born cross
section of $e^{+}e^{-} \rightarrow J/\psi\eta\pi^{0}$ is assumed to
follow a $Y(4260)$ Breit$-$Wigner line shape with resonance parameters
taken from the Particle Data Group (PDG)~\cite{bib23}. Final state
radiation effects associated with charged particles are handled
with {\footnotesize PHOTOS}~\cite{bib24}.

To study possible backgrounds, a MC sample of inclusive $Y(4260)$
decays, equivalent to an integrated luminosity of 825.6 pb$^{-1}$, is
also generated at $\sqrt{s}$ = 4.260 GeV. In these simulations, the
$Y(4260)$ is allowed to decay generically, with the main known decay
channels being generated using {\footnotesize EVTGEN}~\cite{bib25}
with branching fractions set to world average values~\cite{bib23}. The
remaining events associated with charmonium decays are generated with
{\footnotesize LUNDCHARM}~\cite{bib26}, while continuum hadronic events
are generated with {\footnotesize PYTHIA}~\cite{bib27}. QED events
($e^+ e^- \to e^+ e^-$, $\mu^+ \mu^-$, and $\gamma \gamma$) are
generated with {\footnotesize KKMC}~\cite{bib22}. Backgrounds at other
energy points are expected to be similar.

\section{Event selection}
Events with two charged tracks with a net charge of zero are
selected. For each good charged track, the polar angle in the MDC must
satisfy $|\cos \theta|<0.93$, and the point of closest approach to the
$e^{+}e^{-}$ interaction point must be within $\pm$10 cm in the beam
direction and within $\pm$1 cm in the plane perpendicular to the beam
direction. The momenta of leptons from the $J/\psi$
decays in the laboratory frame are required to be larger than 1.0 GeV/$c$.
$E/p$ is used to separate electrons from muons, where $E$ is the energy deposited
in the EMC and $p$ is the momentum measured by
the MDC. For electron candidates, $E/p$ should be larger than 0.7,
while for muons, it should be less than 0.3. To suppress background from events
with pion tracks in the final state, at least one of the two muons is required to
have at least five layers with valid hits in the MUC.

Showers identified as photon candidates must satisfy fiducial and
shower quality as well as timing requirements. The minimum EMC energy
is 25 MeV for barrel showers ($|\cos \theta|<$ 0.80) and 50 MeV for
end cap showers (0.86 $<|\cos \theta|<$ 0.92). To eliminate showers
produced by charged particles, a photon must be separated by at least
5 deg from any charged track. The time information from the EMC is
also used to suppress electronic noise and energy deposits unrelated
to the event. At least four good photon candidates in each event are
required.

To improve the momentum resolution and reduce the background, the
event is subjected to a four-constraint (4C) kinematic fit under the
hypothesis $e^{+}e^{-}\rightarrow \gamma\gamma\gamma\gamma l^{+}l^{-}$
($l$ = $e/\mu$), and the $\chi^{2}$ is required to be less than 40. For
events with more than four photons, the four photons with the smallest
$\chi^{2}$ from the 4C fit are assigned as the photons from $\eta$ and
$\pi^{0}$.

\begin{figure}[H]
  \centering

     \includegraphics[width=0.235\textwidth,height=0.13\textheight]{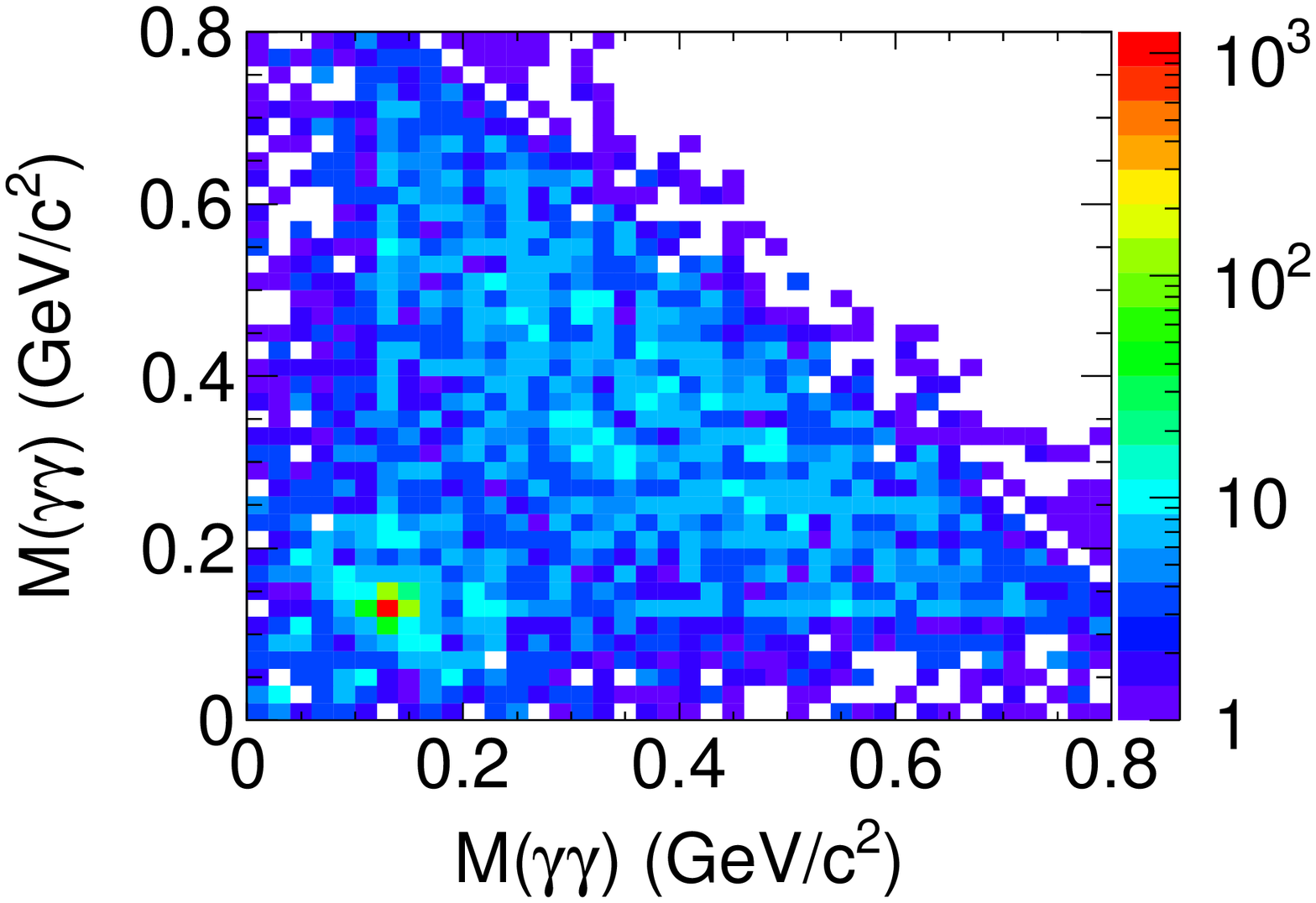}
     \includegraphics[width=0.235\textwidth,height=0.13\textheight]{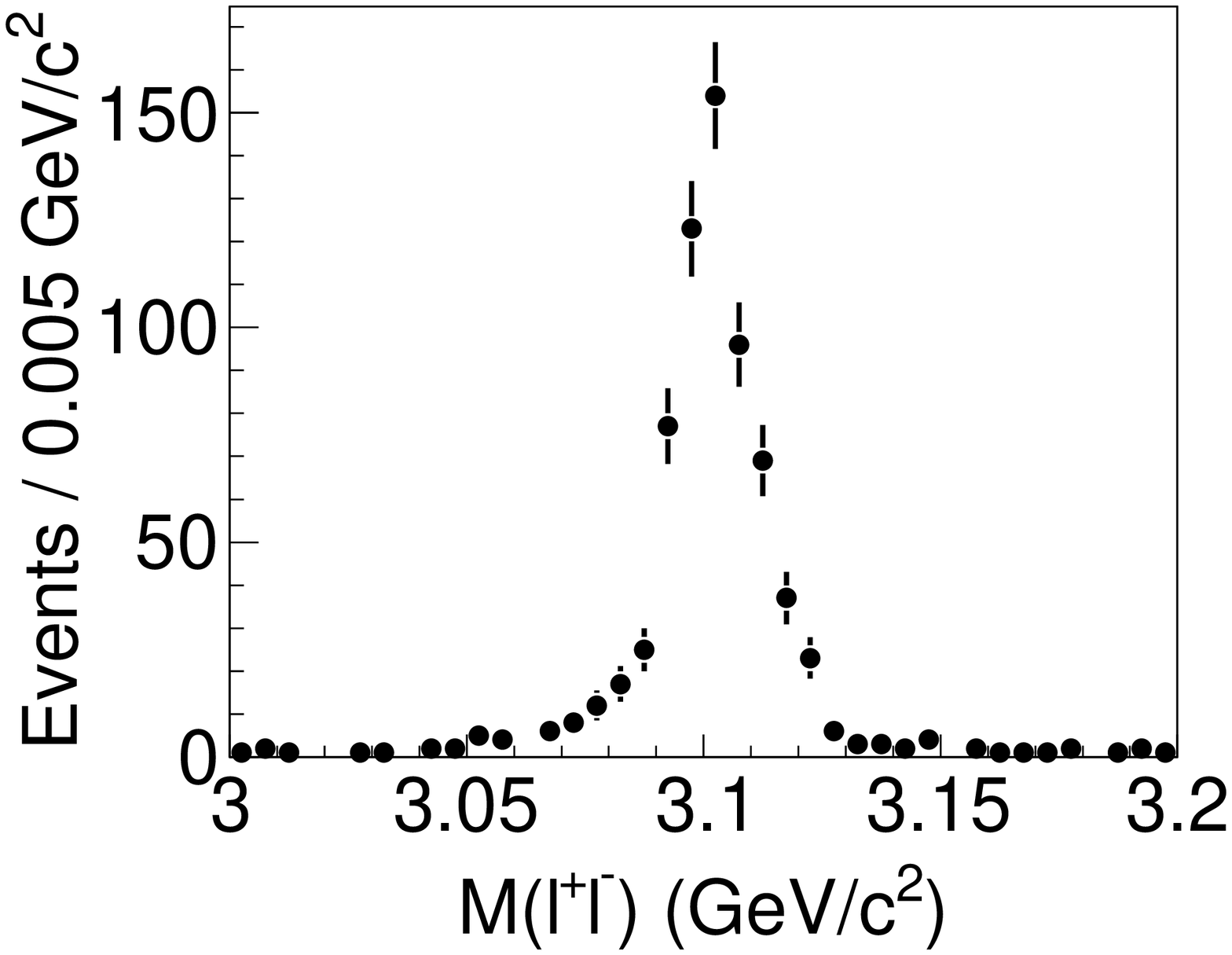}
     \includegraphics[width=0.235\textwidth,height=0.13\textheight]{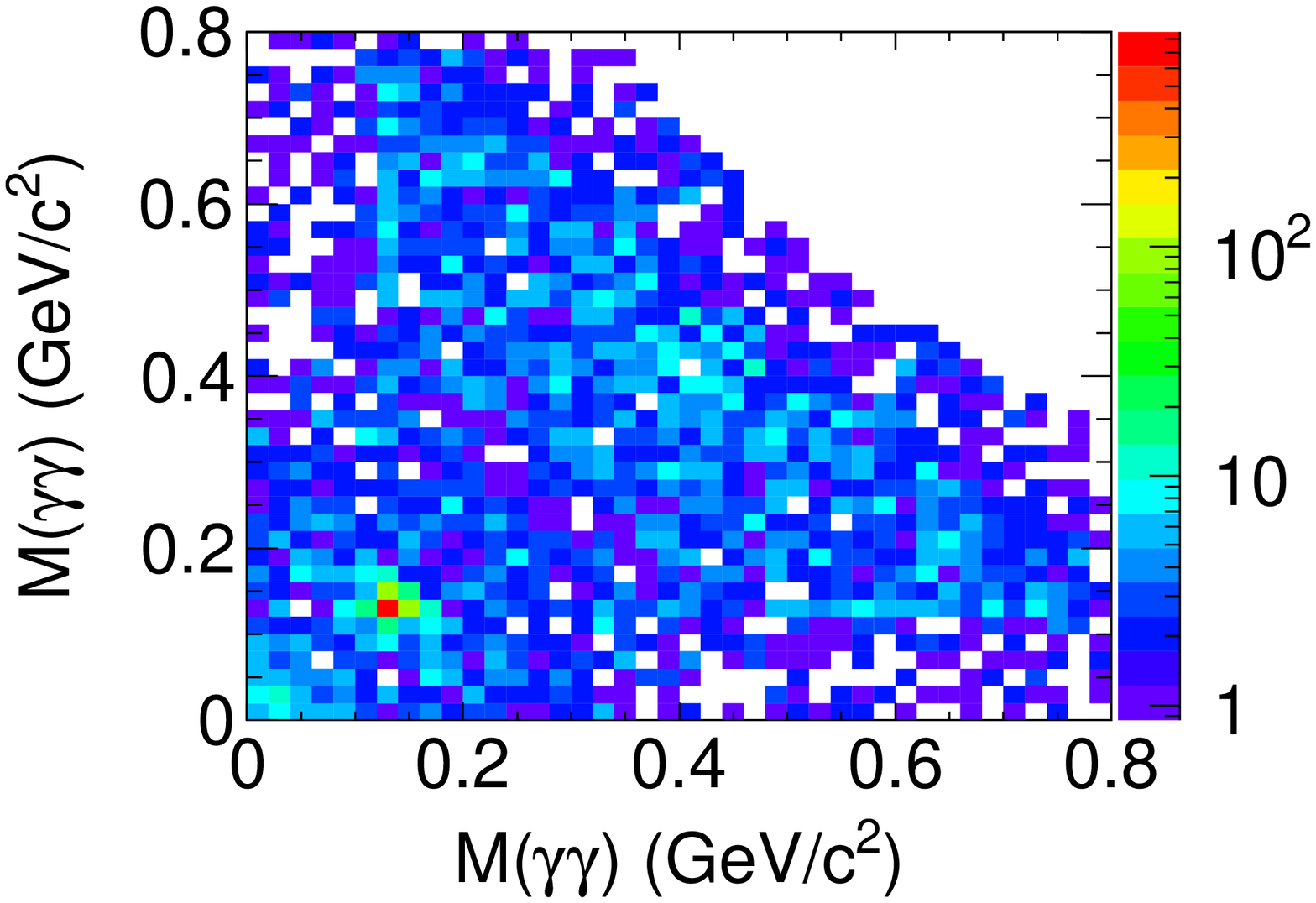}
     \includegraphics[width=0.235\textwidth,height=0.13\textheight]{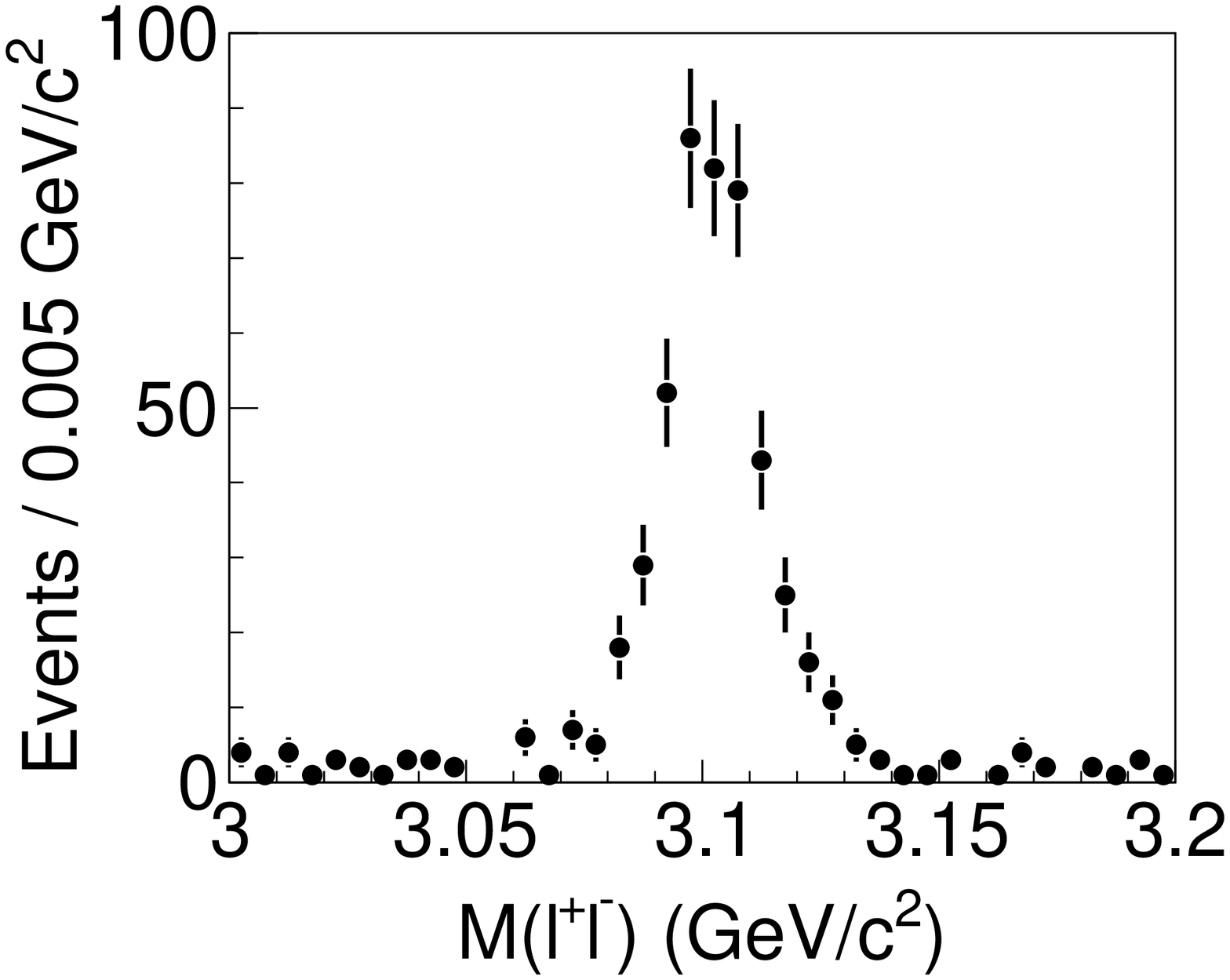}
  \caption{Scatter plot of $M(\gamma\gamma)$ with all six combinations
  for events in the $J/\psi$ signal region (left) and distribution of
  $M(l^{+}l^{-})$ for events in the $\pi^{0}\pi^{0}$ signal region
  (right) for data at $\sqrt{s}$ = 4.226 GeV (top) and 4.257 GeV (bottom). }
 \label{fig:pi0pi0}
\end{figure}

After selecting the $\gamma\gamma\gamma\gamma l^{+}l^{-}$ candidate,
scatter plots of $M(\gamma\gamma)$ with all six combinations of photon
pairs for events in the $J/\psi$ signal region (3.067 $<
M(l^{+}l^{-}) <$ 3.127 GeV/$c^{2}$) for data at $\sqrt{s}$ = 4.226 and 4.257 GeV
 are shown in the left two panels of Fig.~\ref{fig:pi0pi0}. Distributions of
$M(l^{+}l^{-})$ for events in the $\pi^{0}\pi^{0}$ signal region (both
photon pairs satisfy $|M(\gamma\gamma) - m_{\pi^{0}}| < $ 10 MeV/$c^{2}$) for
data at $\sqrt{s}$ = 4.226 and 4.257 GeV are shown in the right two
panels of Fig.~\ref{fig:pi0pi0}. Clear $J/\psi$ peaks are observed, corresponding
to $\pi^{0}\pi^{0}J/\psi$ events. To remove this
$\pi^{0}\pi^{0}J/\psi$ background, events with any combination of
photon pairs in the $\pi^{0} \pi^{0}$ region of the scatter plot are
rejected.

\begin{figure}[H]
  \centering
    \includegraphics[width=0.235\textwidth,height=0.13\textheight]{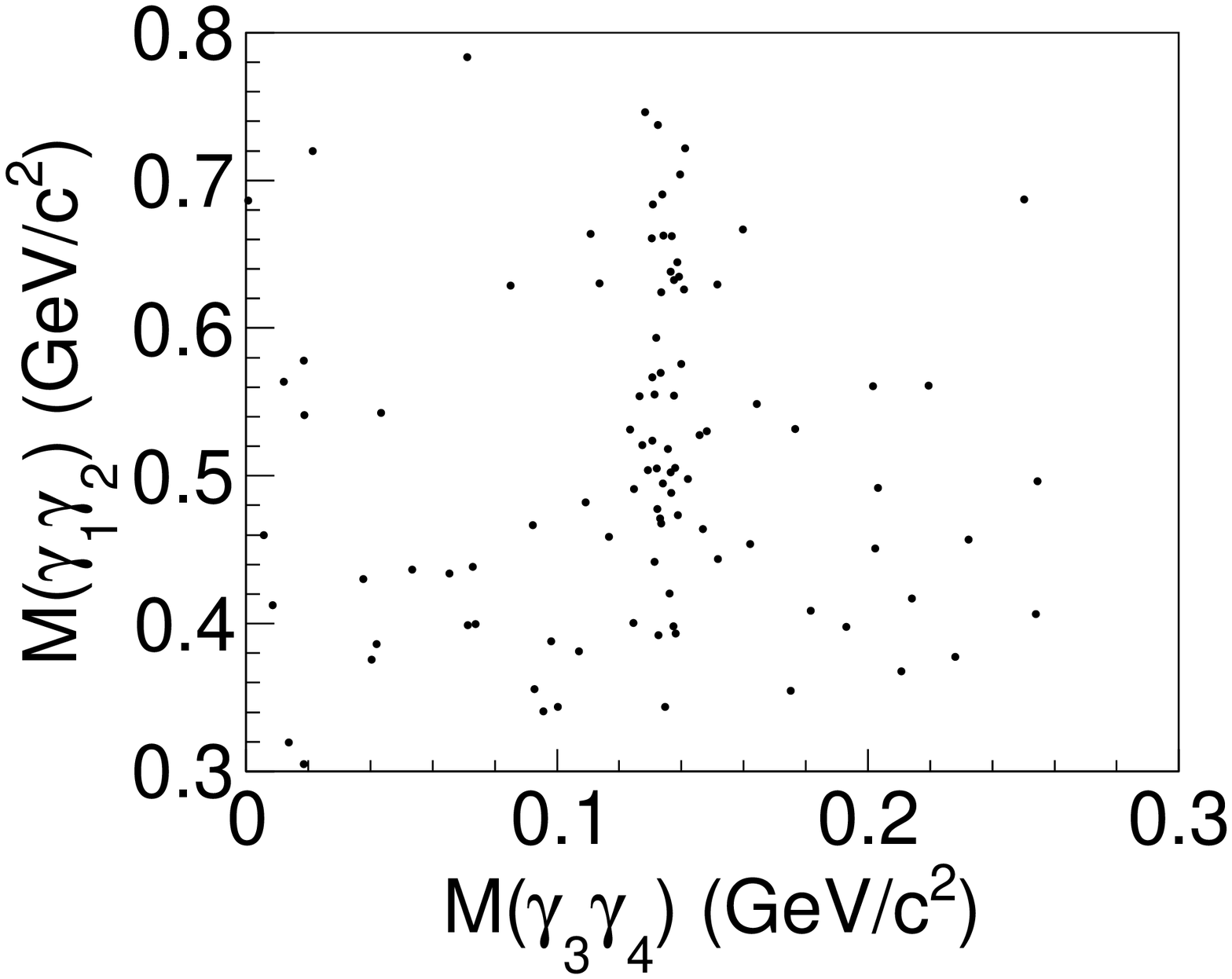}
    \includegraphics[width=0.235\textwidth,height=0.13\textheight]{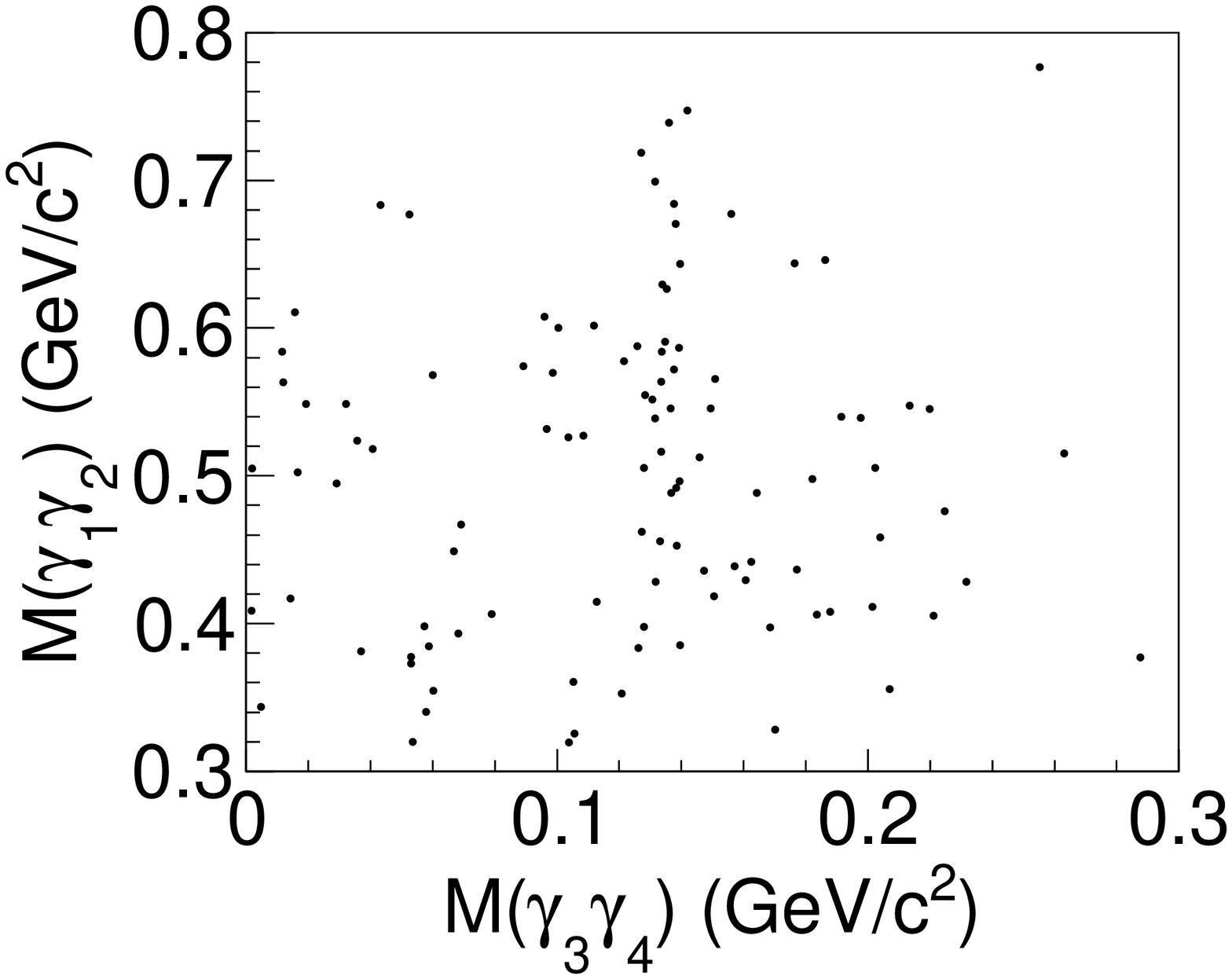}
    \includegraphics[width=0.235\textwidth,height=0.13\textheight]{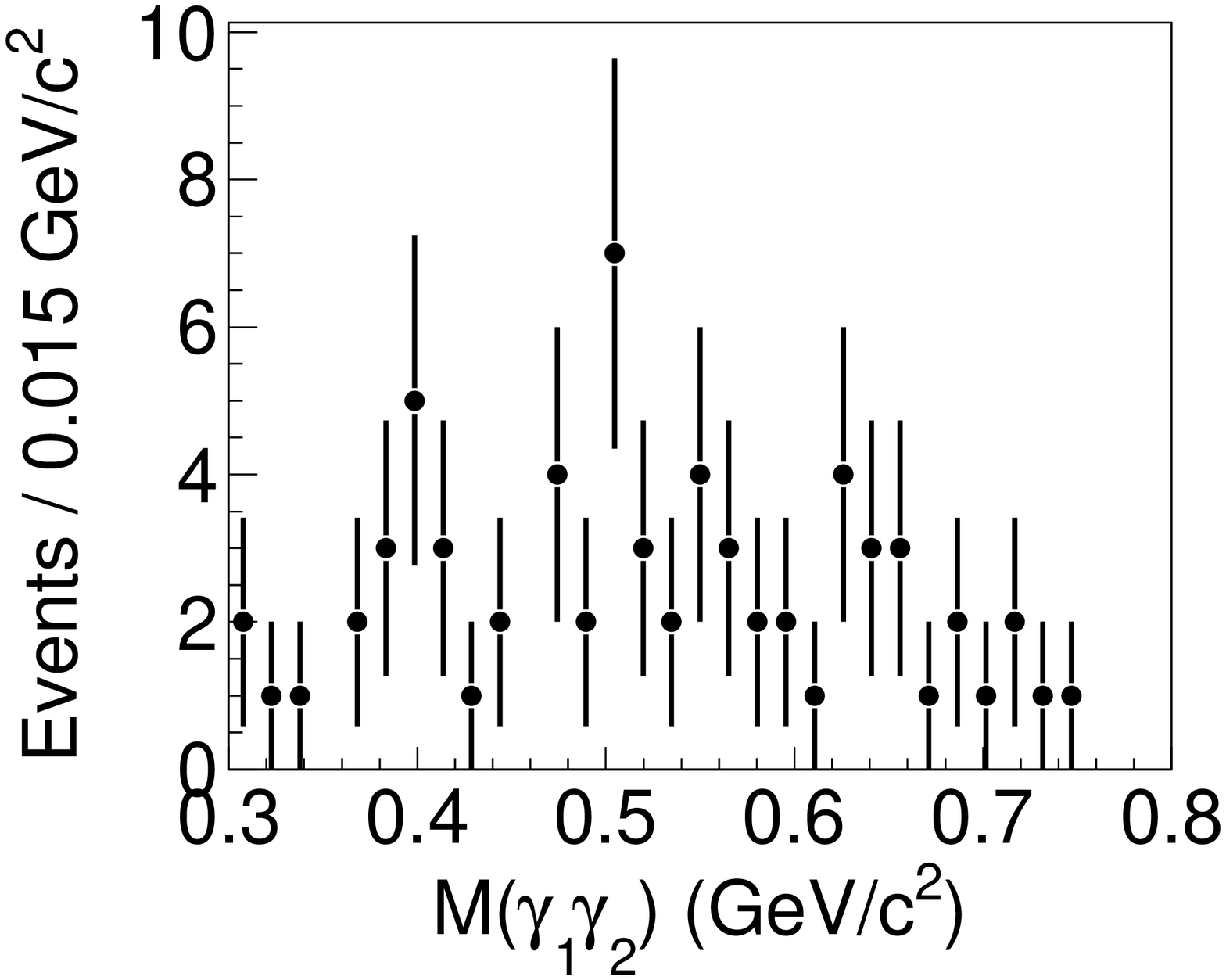}
    \includegraphics[width=0.235\textwidth,height=0.13\textheight]{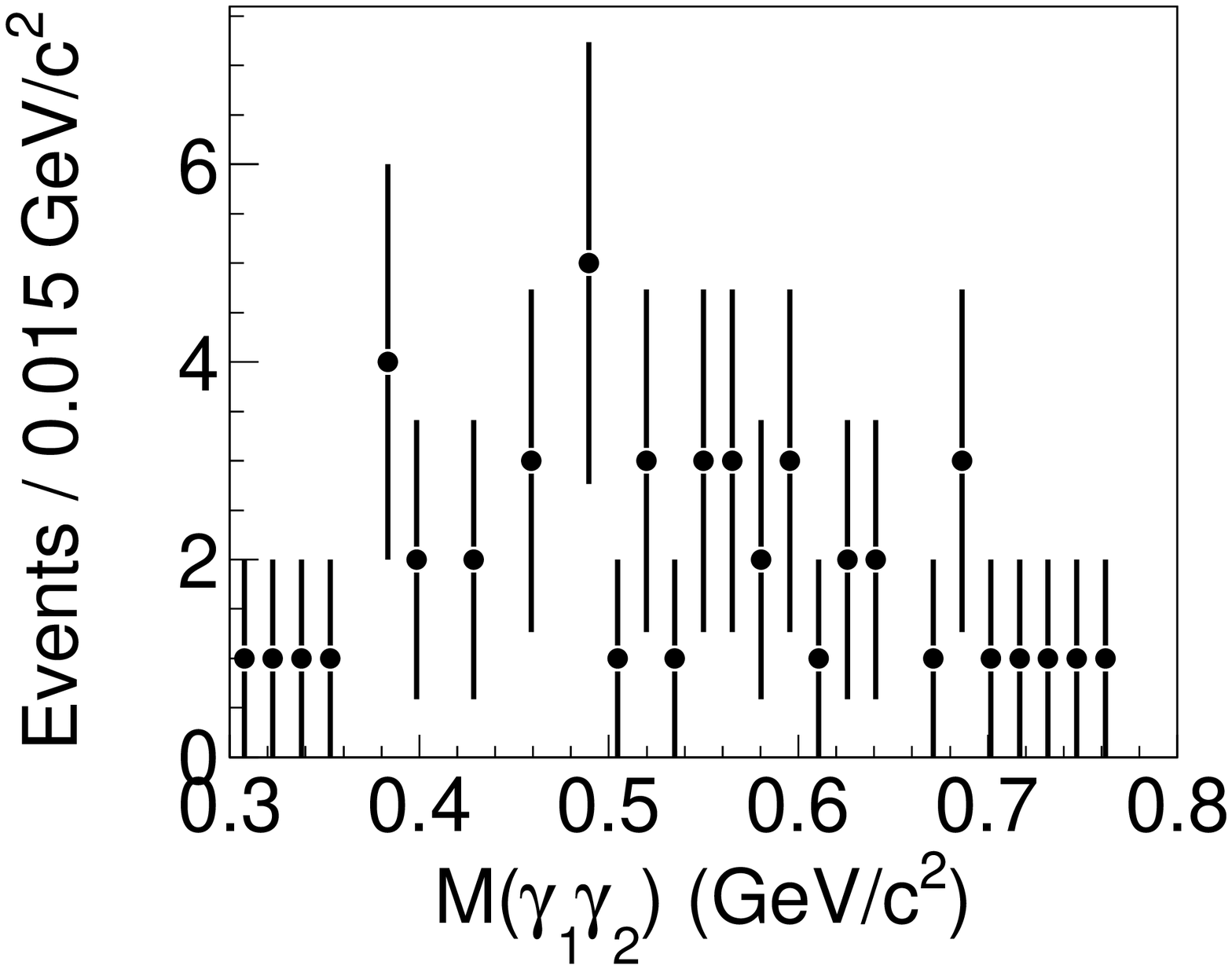}
    \includegraphics[width=0.235\textwidth,height=0.13\textheight]{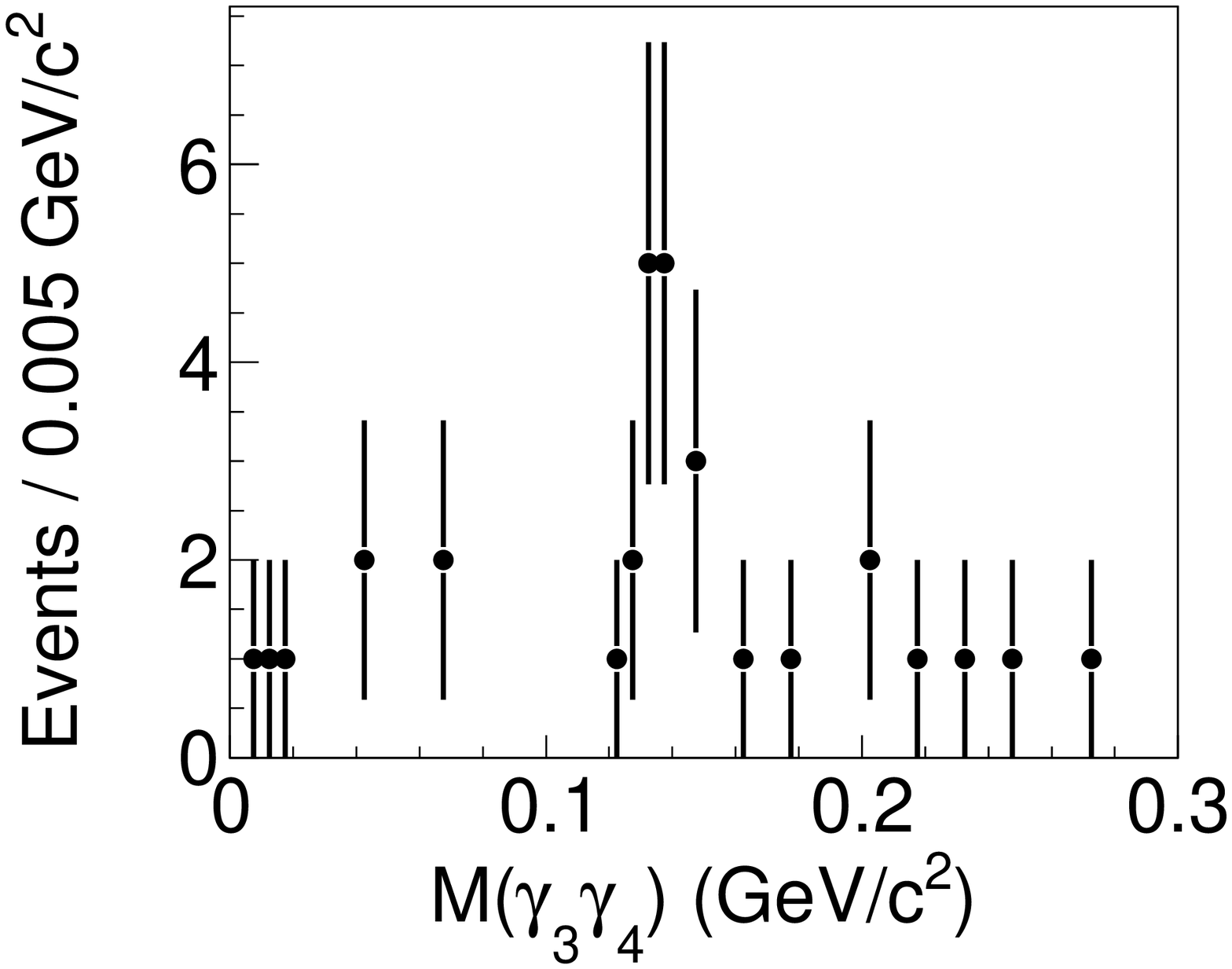}
    \includegraphics[width=0.235\textwidth,height=0.13\textheight]{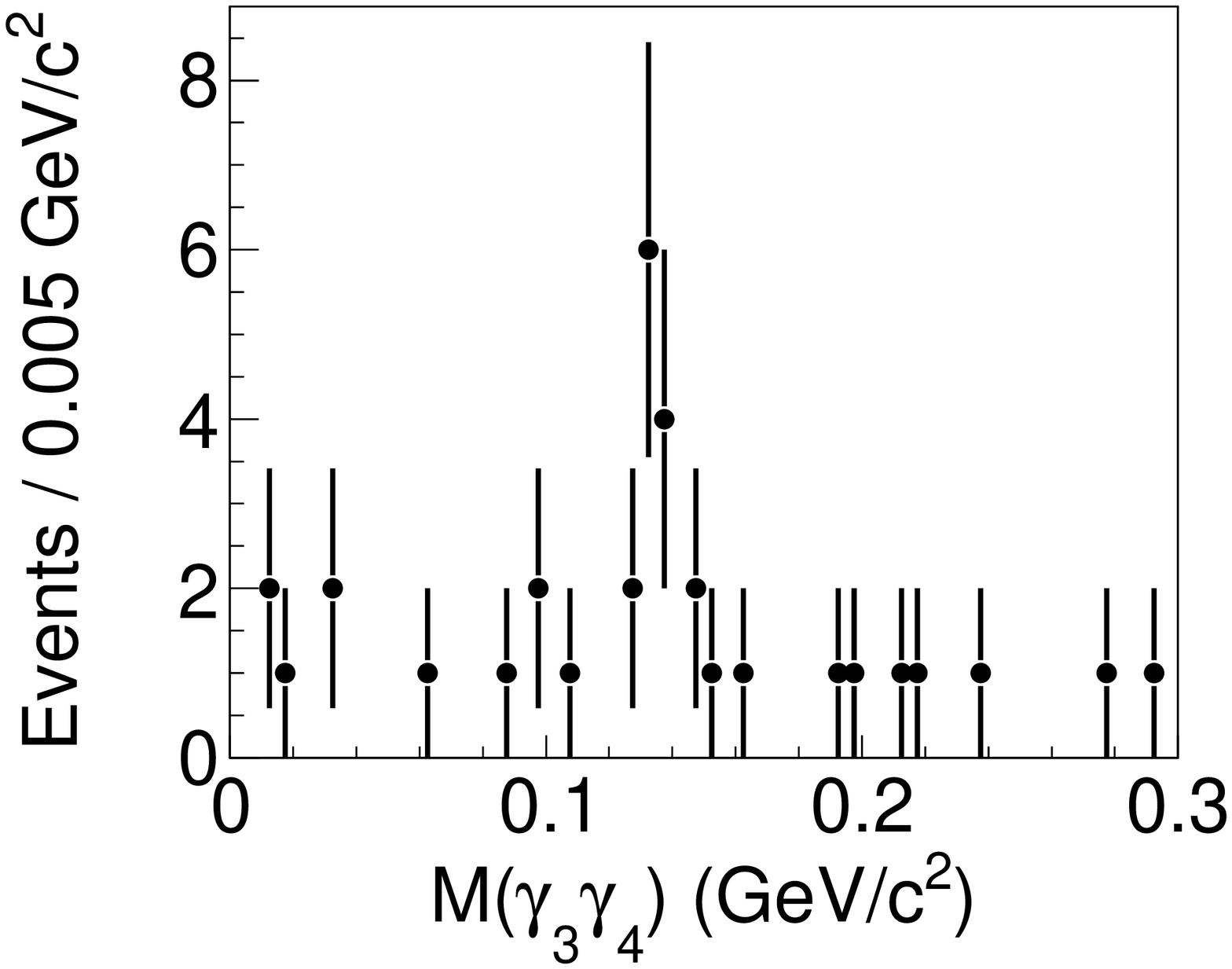}
  \caption{Scatter plot of $M(\gamma\gamma)$ for the combination
  closest to the $\eta\pi^{0}$ signal region for events in the $J/\psi$ signal
  region (top), projection of the scatter plot on $M(\gamma_{1}\gamma_{2})$ with $M(\gamma_{3}\gamma_{4})$ in $\pi^{0}$ signal region (middle), and
  projection of the scatter plot on $M(\gamma_{3}\gamma_{4})$ with $M(\gamma_{1}\gamma_{2})$ in $\eta$ signal region (bottom) for data at
  $\sqrt{s}$ = 4.226 GeV (left) and 4.257 GeV (right).}
  \label{fig:scatter}
\end{figure}

After rejecting the $\pi^{0}\pi^{0}J/\psi$ background, we choose the
combination of photon pairs closest to the $\eta\pi^{0}$ signal
region by minimizing
$\sqrt{|\frac{M(\gamma_{1}\gamma_{2})-m_{\eta}}{\sigma_{\eta}}|^{2}+|\frac{M(\gamma_{3}\gamma_{4})-m_{\pi^{0}}}{\sigma_{\pi^{0}}}|^{2}}$,
where $\sigma_{\eta}$ and $\sigma_{\pi^{0}}$ are the $\eta$ and
$\pi^{0}$ resolutions obtained from the signal MC, respectively. The
scatter plots of $M(\gamma\gamma)$ with the combination closest to the
$\eta\pi^{0}$ signal region for events in the $J/\psi$ signal region
for data at $\sqrt{s}$ = 4.226 and 4.257 GeV are shown in the top two
panels of Fig.~\ref{fig:scatter}. No cluster of $\eta\pi^{0}$ events is observed in the $J/\psi$ signal region, with a vertical band for $\pi^{0}\rightarrow \gamma \gamma$
clearly visible, but no prominent band for $\eta\rightarrow \gamma \gamma$ is observed. The projections of the scatter plots on $M(\gamma_{1}\gamma_{2})$ with $M(\gamma_{3}\gamma_{4})$ in the $\pi^{0}$ signal region ($|M(\gamma_{3}\gamma_{4})-m_{\pi^{0}}|<10$ MeV/$c^{2}$) and projections on $M(\gamma_{3}\gamma_{4})$ with $M(\gamma_{1}\gamma_{2})$ in the $\eta$ signal region ($|M(\gamma_{1}\gamma_{2})-m_{\eta}|<30$ MeV/$c^{2}$) for data are shown in the middle and bottom panels of Fig.~\ref{fig:scatter}, respectively.

\begin{figure}[H]
  \centering
    \includegraphics[width=0.235\textwidth,height=0.13\textheight]{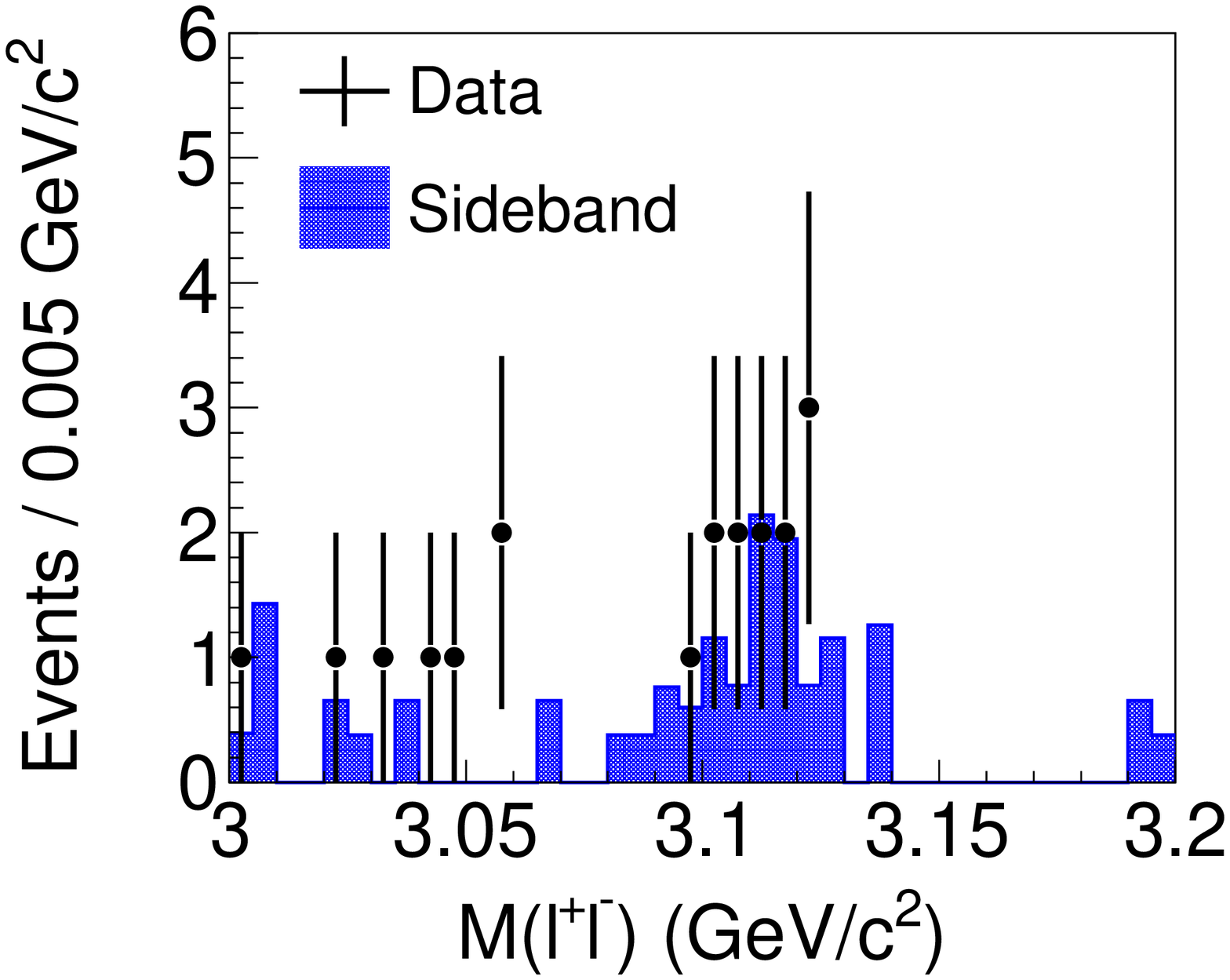}
    \includegraphics[width=0.235\textwidth,height=0.13\textheight]{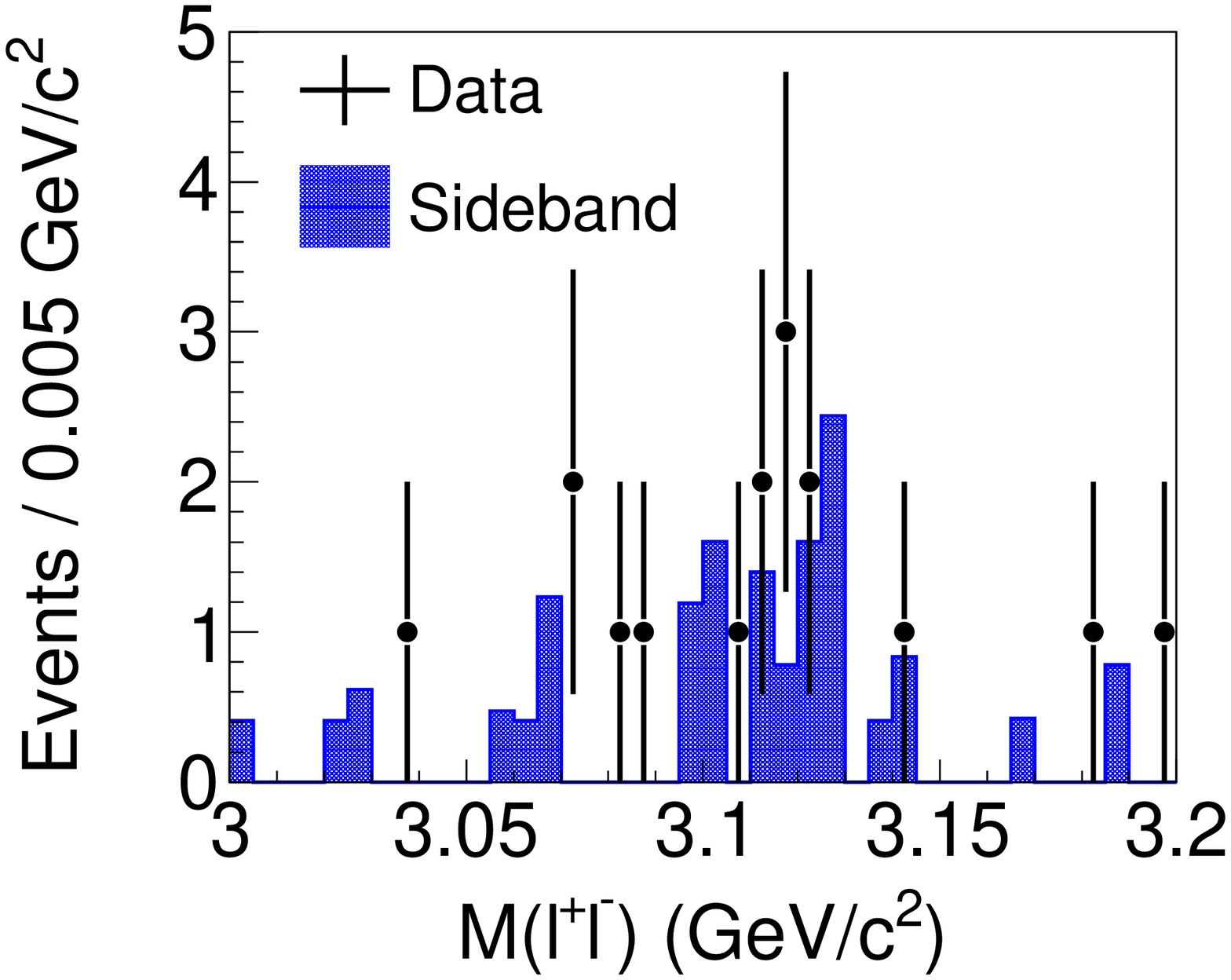}
  \caption{Distributions of $M(l^{+}l^{-})$ for events in
  $\eta\pi^{0}$ signal region and sideband regions for data at $\sqrt{s}$ = 4.226 GeV (left) and 4.257
  GeV (right). The error bars are the $M(l^{+}l^{-})$ distributions for events in the
  $\eta\pi^{0}$ signal region, and the shaded histograms are those in
  the $\eta\pi^{0}$ sideband regions.}
  \label{fig:etapi0}
\end{figure}

The background for $e^{+}e^{-}\rightarrow J/\psi \eta \pi^{0}$
is studied using the inclusive MC sample at $\sqrt{s}$ =
4.260 GeV. After imposing all event selection requirements, there are two background events from $e^{+}e^{-}\rightarrow
\pi^{0}\pi^{0}J/\psi$ and nine background events arising from $e^{+}e^{-}\rightarrow
\gamma_\text{ISR}\psi', ~\gamma_\text{ISR}\psi''$, and
$\gamma_\text{ISR}\psi(4040)$. No other background survives. The
background can be evaluated with $\eta \pi^{0}$ sideband
events. Distributions of $M(l^{+}l^{-})$ for events in the $\eta\pi^{0}$
signal region for data at $\sqrt{s}$ = 4.226
and 4.257 GeV are shown in Fig.~\ref{fig:etapi0}. Distributions of $M(l^{+}l^{-})$ for events corresponding to the
normalized two-dimensional $\eta\pi^{0}$ sidebands are shown as shaded
histograms. The $\eta$ sideband regions are defined as 0.3978 $< M(\gamma_{1}\gamma_{2})<$ 0.4578 GeV/$c^{2}$
and 0.6378 $< M(\gamma_{1}\gamma_{2})<$ 0.6978 GeV/$c^{2}$. The $\pi^{0}$ sideband
regions are defined as 0.0849 $<
M(\gamma_{3}\gamma_{4}) < $ 0.1049 GeV/$c^{2}$ and 0.1649 $<
M(\gamma_{3}\gamma_{4}) < $ 0.1849 GeV/$c^{2}$. The counted number of observed events
in the $J/\psi \eta \pi^{0}$ signal region $N^\text{obs}$ and number of
background events estimated from $\eta \pi^{0}$ sidebands $N^\text{bkg}$ are listed
in Table~\ref{tab:results}.

\section{Cross section upper limits}
  Since no $J/\psi\eta\pi^{0}$ signal above the background is observed,
  upper limits on the Born cross section of $e^{+}e^{-}\rightarrow J/\psi\eta\pi^{0}$ at the 90\% C.L.
  are determined using the formula
\begin{eqnarray}
\sigma^\text{Born}<\frac{N^\text{up}_\text{observed}}{{\cal
  L}(1+\delta^{r})(1+\delta^{v})(\epsilon^{ee}{\cal B}^{ee} + \epsilon^{\mu\mu}{\cal
  B}^{\mu\mu}){\cal B}^{\pi^{0}}{\cal B}^{\eta}}~,
\label{eq:xsec}
\end{eqnarray}
where $N^\text{up}_\text{observed}$ is the upper limit on the number of signal
events; $\cal{L}$ is the integrated luminosity; $(1+\delta^{r})$ is
the radiative correction factor, which is taken from a QED
calculation assuming the $e^{+}e^{-}\rightarrow J/\psi\eta\pi^{0}$
cross section is described by a $Y(4260)$ Breit$-$Wigner line shape with
parameters taken from the PDG~\cite{bib23}; $(1+\delta^{v})$ is the
vacuum polarization factor including leptonic and hadronic parts and
taken from a QED calculation with an accuracy of 0.5\%~\cite{bib28};
$\epsilon^{ee}$ and $\epsilon^{\mu\mu}$ are
the efficiencies for $e^{+}e^{-}$ and $\mu^{+}\mu^{-}$ modes,
respectively; ${\cal B}^{ee}$ and ${\cal B}^{\mu\mu}$ are the branching
fractions of $J/\psi\rightarrow e^{+}e^{-}$ and $J/\psi\rightarrow
\mu^{+}\mu^{-}$~\cite{bib23}, respectively; and ${\cal B}^{\eta}$ and
${\cal B}^{\pi^{0}}$ are the branching fractions of $\eta\rightarrow
\gamma\gamma$ and $\pi^{0}\rightarrow \gamma\gamma$~\cite{bib23},
respectively.

 The efficiency corrected upper limit on the number of signal events
 $N^\text{up} \equiv \frac{N^\text{up}_\text{observed}}{\epsilon^{ee}{\cal B}^{ee} +
 \epsilon^{\mu\mu}{\cal B}^{\mu\mu}}$ is estimated with $N^\text{obs}$ and
 $N^\text{bkg}$ using the profile likelihood method, which is
 implemented by {\footnotesize TRolke} in the {\footnotesize ROOT}
 framework~\cite{bib29}. The calculation for obtaining $N^{up}$
 includes the background fluctuation and the systematic uncertainty of
 the cross section measurement. The background fluctuation is assumed
 to follow a Poisson distribution. The systematic uncertainty of the
 cross section is taken as a Gaussian uncertainty.

 The systematic uncertainty of the cross section measurement in
 Eq.~(\ref{eq:xsec}) includes the luminosity measurement, detection efficiency, and 
 intermediate decay branching fractions. The systematic
 uncertainties of the luminosity, track reconstruction, and photon
 detection are 1.0\%~\cite{bib11}, 1.0\% per track~\cite{bib30}, and
 1.0\% per photon~\cite{bib31}, respectively. The systematic
 uncertainties from the branching fraction of $\pi^{0}$ and $\eta$ decays
 are taken from the PDG~\cite {bib23}. These sources of systematic uncertainty,
 which are summarized in the top part of Table~\ref{tab:sys}, are common for
 $e^{+}e^{-}$ and $\mu^{+}\mu^{-}$ modes. The following sources of systematic
 uncertainty, which are uncorrelated for the $e^{+}e^{-}$ and
 $\mu^{+}\mu^{-}$ modes, are summarized in the bottom part of Table~\ref{tab:sys}.
 The systematic uncertainty from the branching fraction of $J/\psi$
 decay is taken from the PDG~\cite {bib23}. The systematic uncertainty from
 the requirement on the number of MUC hits is 3.6\% and estimated by comparing the efficiency
 of the MUC requirement between data and MC in the control sample $e^{+}e^{-}\rightarrow \pi^{0}\pi^{0}J/\psi$
 at $\sqrt{s}$ = 4.257 GeV. The systematic uncertainty from the requirement of the $J/\psi$
 signal region is estimated by smearing the invariant mass of $l^{+}l^{-}$ of the signal MC
 with a Gaussian function to compensate for the resolution difference between the data and
 MC when calculating the efficiency. The parameters for smearing are determined by fitting
 the $J/\psi$ distribution of data with the MC shape convoluted with a Gaussian function for
 the control sample $e^{+}e^{-}\rightarrow \pi^{0}\pi^{0}J/\psi$. The difference in the
 detection efficiency between signal MC samples with and without the smearing is taken as the systematic
 uncertainty. The systematic uncertainty from the MC model is estimated by generating a MC
 sample with the angular distribution of leptons determined from the 
 $\pi^{+}\pi^{-}J/\psi$ data. The systematic uncertainty due to kinematic
 fitting is estimated by correcting the helix parameters of charged
 tracks according the method described in Ref.~\cite{bib32}, where the
 correction factors are obtained from the control sample $\psi'\rightarrow \gamma\chi_{cJ}$
 and the difference in the detection efficiency between with and
 without making the correction to the MC is taken as the systematic uncertainty.
 The uncorrelated systematic uncertainties for the electron and muon channels are combined by taking
 the weighted  average with weights $\epsilon^{ee}\mathcal{B}^{ee}$ and
 $\epsilon^{\mu\mu}\mathcal{B}^{\mu\mu}$, respectively.
 The total systematic uncertainty is obtained by summing all the sources of
 the systematic uncertainty in quadrature.

\begin{table*}[!htbp]
\centering \caption{\label{tab:results}
Results on $e^{+}e^{-}\rightarrow
J/\psi\eta\pi^{0}$. Listed in the table are the integrated luminosity
$\cal{L}$, radiative correction factor (1+$\delta^{r}$) taken from
QED calculation assuming the $Y(4260)$ cross section follows a
Breit$-$Wigner line shape, vacuum polarization factor (1+$\delta^{v}$),
average efficiency ($\epsilon^{ee}{\cal B}^{ee}$ +
$\epsilon^{\mu\mu}{\cal B}^{\mu\mu}$), number of observed events $N^\text{obs}$,
number of estimated background events $N^\text{bkg}$, the efficiency corrected
upper limits on the number of signal events $N^\text{up}$, and upper limits on the
Born cross section $\sigma^\text{Born}_\text{UL}$ (at the 90 \% C.L.)
at each energy point. }
\begin{tabular}{lcccccccc}
\hline
$\sqrt{s}$ (GeV)    &$\cal{L}$ (pb$^{-1}$)  &(1+$\delta^{r}$) &(1+$\delta^{v}$) &($\epsilon^{ee}{\cal B}^{ee}$ + $\epsilon^{\mu\mu}{\cal B}^{\mu\mu}$) (\%)  &$N^\text{obs}$  &$N^\text{bkg}$  &$N^\text{up}$   &$\sigma^\text{Born}_\text{UL}$ (pb) \\
\hline
4.009            &482.0    &0.838  &1.044    &$2.1 \pm 0.1(sys.)$    &5    &1   &598.1     &3.6 \\
4.226            &1047.3   &0.844  &1.056    &$2.2 \pm 0.1(sys.)$    &12   &11  &592.9     &1.7\\
4.257            &825.6    &0.847  &1.054    &$2.2 \pm 0.1(sys.)$    &12   &8   &654.1     &2.4\\
4.358            &539.8    &0.942  &1.051    &$2.2 \pm 0.1(sys.)$    &5    &4   &283.2     &1.4\\
4.416            &1028.9   &0.951  &1.053    &$2.3 \pm 0.1(sys.)$    &5    &6   &342.7     &0.9\\
4.599            &566.9    &0.965  &1.055    &$2.4 \pm 0.1(sys.)$    &6    &3   &418.4     &1.9\\
\hline
\end{tabular}
\end{table*}

\begin{table*}[!htbp]
\centering
\caption{\label{tab:sys} Systematic uncertainties in the $J/\psi \eta
\pi^{0}$ cross section measurement at each energy point (in \%). The
items in parentheses in the bottom part of the table are the
uncorrelated systematic uncertainties for the $e^{+}e^{-}$ (first) and
$\mu^{+}\mu^{-}$ (second) modes. }
\begin{tabular}{lcccccc}
\hline
Sources/$\sqrt{s}$ (GeV)    &4.009  &4.226  &4.257   &4.358  &4.416 &4.599 \\
\hline
 Luminosity              &1.0     &1.0   &1.0   &1.0  &1.0  &1.0 \\

 MDC tracking            &2.0     &2.0   &2.0   &2.0 &2.0   &2.0 \\

 Photon reconstruction   &4.0     &4.0   &4.0   &4.0 &4.0   &4.0\\

 ${\cal B}(\pi^{0}\rightarrow \gamma\gamma$), {\cal B}($\eta\rightarrow
 \gamma\gamma$)                           &0.5 &0.5 &0.5    &0.5 &0.5    &0.5 \\ \hline
 ${\cal B}(J/\psi\rightarrow l^{+}l^{-}$) &(0.5, 0.5)    &(0.5, 0.5)   &(0.5, 0.5)   &(0.5, 0.5)   &(0.5, 0.5)   &(0.5, 0.5) \\
 MUC hits                                 &(0, 3.6)      &(0, 3.6)     &(0, 3.6)     &(0, 3.6)     &(0, 3.6)     &(0, 3.6)  \\
 $J/\psi$ mass resolution                 &(0.2, 1.3)    &(0.8, 1.2)   &(0.5, 1.3)   &(0.2, 0.7)   &(0.7, 1.6)   &(0.1, 0.6)\\
 Decay model                              &(1.5, 1.9)    &(0.9, 1.1)   &(0.4, 0.6)   &(0.2, 0.7)   &(0.7, 0.2)   &(0.2, 0.2)\\
 Kinematic fitting                        &(1.2, 0.9)    &(1.1, 1.2)   &(0.9, 0.9)   &(0.7, 1.2)   &(1.1, 1.0)   &(1.0, 1.4)  \\
 \hline
 Total                                    &5.3           &5.3          &5.2          &5.2          &5.3          &5.2\\

\hline
\end{tabular}
\end{table*}

The systematic uncertainty on the size of the background is estimated
by evaluating $N^\text{up}$ with different signal and sideband regions for
$\eta$ and $\pi^{0}$. The most conservative $N^\text{up}$ is taken as the
final result, as listed in Table~\ref{tab:results}.
The upper limits on the Born cross section of $e^{+}e^{-}\rightarrow
J/\psi\eta\pi^{0}$ ($\sigma^\text{Born}_\text{UL}$) assuming it follows a $Y(4260)$
Breit$-$Wigner line shape are listed in Table~\ref{tab:results}.

For comparison, the radiative correction factor and detection efficiency have been recalculated
assuming the $e^{+}e^{-}\rightarrow J/\psi\eta\pi^{0}$ cross section
follows alternative line shapes. If the cross section follows the line shape
of the $Y(4040)$, the upper limit on the Born cross section is 4.1 pb at
$\sqrt{s}$ = 4.009 GeV. For a $Y(4360)$ line shape, it is 1.6 pb at
$\sqrt{s}$ = 4.358 GeV. For a $Y(4415)$ line shape, it is 1.5 pb at
$\sqrt{s}$ = 4.358 GeV and 1.0 pb at $\sqrt{s}$ = 4.416 GeV. For a
$Y(4660)$ line shape, it is 2.0 pb at $\sqrt{s}$ = 4.599 GeV.

It is also possible to set upper limits on $e^{+}e^{-}\rightarrow
Z^{0}_{c}\pi^{0}\rightarrow J/\psi \eta \pi^{0}$. The number of
observed events and number of estimated background events in the $Z^{0}_{c}$ signal region ($3.850  <
M(J/\psi \eta) < 3.940\;\text{GeV}/c^{2}$) are 7 and 4 $\pm 2$, respectively, at $\sqrt{s}$ = 4.226
GeV, and 8 and $3 \pm 2$, respectively, at $\sqrt{s}$ = 4.257 GeV. The upper limit on $\sigma
(e^{+}e^{-}\rightarrow Z^{0}_{c}\pi^{0}\rightarrow J/\psi \eta
\pi^{0}$) is determined to be 1.3 pb at $\sqrt{s}$ = 4.226 GeV and
2.0 pb at $\sqrt{s}$ = 4.257 GeV, where only the statistical
uncertainty is given. Compared to the measured cross section of $e^{+}e^{-}\rightarrow Z^{0}_{c}\pi^{0}\rightarrow J/\psi \pi^{0}
\pi^{0}$~\cite{bib33}, the upper limit on the ratio of
the branching fraction $\frac{{\cal B}(Z^{0}_{c}\rightarrow J/\psi \eta)}{{\cal B}(Z^{0}_{c}\rightarrow J/\psi \pi^{0})}$ at the 90\% confidence
level is
0.15 at $\sqrt{s}$ = 4.226 GeV and
0.65 at $\sqrt{s}$ = 4.257 GeV.

\section{Summary}
In summary, using data collected with the BESIII detector, a search
for the isospin violating decay $Y(4260)\rightarrow J/\psi \eta
\pi^{0}$ is performed. No statistically significant signal is observed. The Born cross sections of
$e^{+}e^{-}\rightarrow J/\psi \eta \pi^{0}$ at the 90\% confidence
level limits at $\sqrt{s}$ = 4.009, 4.226, 4.257, 4.358, 4.416, and 4.599 GeV are determined to
be 3.6, 1.7, 2.4, 1.4, 0.9, and 1.9 pb,
respectively. The upper limits are well above the prediction for the
molecule model~\cite{bib18}.

 \section*{Acknowledgement}
The BESIII Collaboration thanks the staff of BEPCII and the IHEP computing center for their strong support. This work is supported in part by National Key Basic Research Program of China under Contract No. 2015CB856700; National Natural Science Foundation of China (NSFC) under Contracts No. 11125525, No. 11235011, No. 11322544, No. 11335008, and No. 11425524; the Chinese Academy of Sciences (CAS) Large-Scale Scientific Facility Program; Joint Large-Scale Scientific Facility Funds of the NSFC and CAS under Contracts No. 11179007, No. U1232201, and No. U1332201; CAS under Contracts No. KJCX2-YW-N29 and No. KJCX2-YW-N45; 100 Talents Program of CAS; INPAC and Shanghai Key Laboratory for Particle Physics and Cosmology; German Research Foundation DFG under Contract No. Collaborative Research Center CRC-1044; Istituto Nazionale di Fisica Nucleare, Italy; Ministry of Development of Turkey under Contract No. DPT2006K-120470; Russian Foundation for Basic Research under Contract No. 14-07-91152; US Department of Energy under Contracts No. DE-FG02-04ER41291, No. DE-FG02-05ER41374, No. DE-FG02-94ER40823, and No. DESC0010118; US National Science Foundation; University of Groningen and the Helmholtzzentrum fuer Schwerionenforschung GmbH, Darmstadt; and the WCU Program of National Research Foundation of Korea under Contract No. R32-2008-000-10155-0.

\end{document}